# Analyzing Mass School Shootings in the United States from 1999 to 2024 with Game Theory, Probability Analysis, and Machine Learning

Wei Dai*,  Senior Member, IEEE, Rui Zhang†, Diya Kafle‡

Department of Computer Science

*Purdue University Northwest*

*Hammond, Indiana, USA*

{weidai*,zhan5097†, kafled‡ }@pnw.edu

*Abstract*

Public safety is vital to every country, especially school safety. In the United States, students and educators are concerned about school shootings. There are critical needs to understand the patterns of school shootings. Without this understanding, we cannot take action to prevent school shootings. Existing research that includes statistical analysis usually focuses on public mass shootings or just shooting incidents that have occurred in the past and there are hardly any articles focusing on mass school shootings. Here we firstly define mathematic models through gam theory. Then, we evaluate shootings events in schools for recently 26-year (1999-2024). Compared with the number of mass school shootings in COVID-19 period, we predict the number of mass school shooting events in the US will be reduced through four machine learning models. We also identify that mass school shootings usually take average 31 minutes with four periods. The annual probability of mass school shootings is $1.23 \times 10^{-5}$ (or one in 81,604) per school. The shootings mostly occur inside buildings, especially classrooms and hallways. By interpreting these data and conducting various statistical analysis, this will ultimately help the law enforcement and schools to reduce the future school shootings. The research data sets could be downloaded via the website: https://publicsafetyinfo.com

## I. INTRODUCTION

Gun violence in schools cause widespread attention across the US. Students, faculty, and staff on campuses could be involved with these shootings, as victims, perpetrators, etc.[1]. These gun-related crimes jeopardize school safety. As of September 19, 2024, students and educators have experienced 50 more school shootings in the United States. There have been resulted 24 deaths and 66 injuries at 50 schools, including 13 colleges and 37 K-12 schools [2]. School shootings include rampage shootings, mass murders, targeted shootings, terrorist attacks, and government shootings [3]. By FBI definition, a mass shooting is defined as any event where four or more people are shot with a gun, but not counting the perpetrator. However, the definition of mass shooting could be "at least three victims" before 2014 [4] [5] [6]. An increase in the number of victims in a mass shooting indicates that the standard for mass shootings has been changed. In this study, we define a mass school shooting as four or more victims are shot with a firearm but does not include the active shooter(s); the incident must occur on school property during the school days or in the immediate period preceding or succeeding it.

The major contributions of the research have been summarized as follows: (1) We designed mathematical models including active shooters and police officers based on game theory. (2) After analyzing school shootings and mass school shootings that are widely used in research papers, we conduct an annual probability of mass school shooting per school.(3) We built four machine learning models to predict the trend of mass school shootings from 2025 to 2030 in the US. (4) This paper critically examines the timeline of mass school shootings. Our focus is mainly on elucidating the four stages of mass school shootings. (5) The paper proposes a statistical analysis to determine the correlation relationship between casualties and five factors including police/hospital distances.

The remainder of the paper is organized as follows. Section II is the related work. Section III is the game theory for the shooting events. Section IV is the probability of Shooting events in schools. Section V is the analysis of mass school shootings. Section VI provides case studies. Section VIII is the conclusion.

## II. RELATED WORK

The trend of school shooting incidents is increased. In [7], authors identified that the incidences of school shootings are related to multiple factors: the incidence of school shootings was found to be higher in states that have a lower proportion of

The Provost Office of Purdue University Northwest and the Indiana Space Grant Consortium funded the research project.

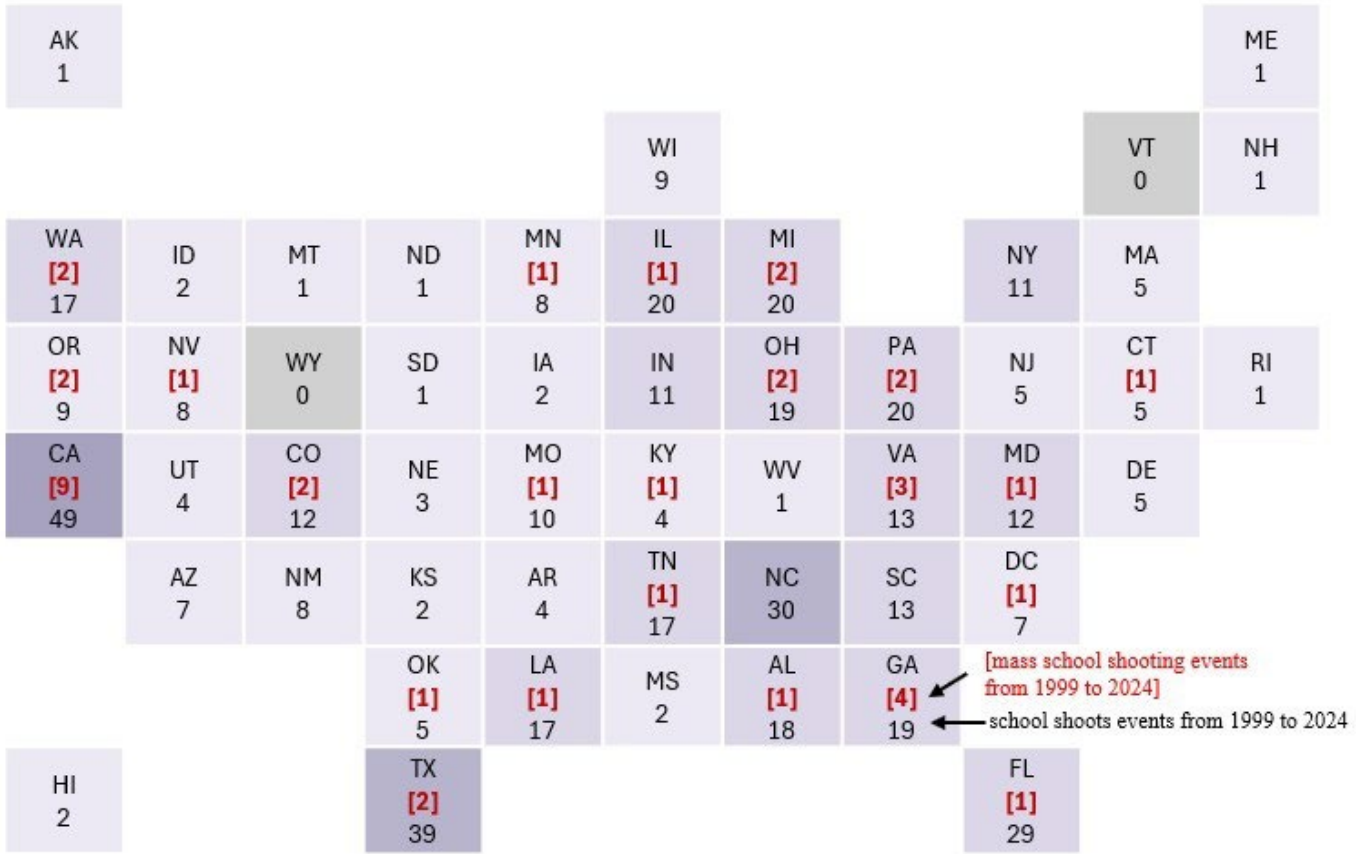

Fig. 1. The number of School Shootings and Mass school shootings by state in the US from 1999 to 2024

their population living in urban areas, do not require background checks for gun and bullet purchases, and exist lower levels of expenditure on mental health and PreK-12 education.

The existing research activities have been understood why the rate of mass shootings in the United States is increasing. A mass shooting is a complex phenomenon with a multitude of contributing factors. Perpetrators have experimented serious mental illness [8], inequality [9], and uncontrolled resentment [10]. It is not the case that serious mental illness (such as schizophrenia, bipolar disorder, and major depression) is a sufficient condition for mass violence [8].

A period of school shootings is essential. The faster active shooters are stopped, the better. In [11], authors note that 47 percent of school shootings lasted less than 15 minutes from the first gunshot to the time the active shooter(s) was terminated; 25 percent of school shootings continued for five minutes. However, the research does not provide detailed time periods for mass school shootings.

## III. GAME THEORY MODELS FOR SHOOTINGS

Game theory is the study of mathematical models in multiple scientific fields, including computer science, social science, and economics. Existing research shows that game theory could be used for homeland security [12] [13] [14]. In this study, mathematical models include active shooters and law enforcements that define players. Active shooters, $i \in n$ intentionally maximize human casualties, $Loss_v(.)$, and material damage, $Loss_m(.)$ in a specific timeline, $t$. Obviously, killers have two restrictions: the total number of bullets, $b \in B$, and the total criminal time, $t \in T$. For the strategy formula of active killers, see Equ. 1. The victimized injury scale, $I(t)$, is extremely high from $t_{attack}$ to $t_{cop}$ because people are not ready for mass shooting. After law enforcement officers arrive at the place to protect victims, the victimized injury scale, $I(t)$, is reduced. The relationship between human casualties, $Loss_v(.)$, and the injury scale of victims, $I(t)$, is defined as Miller Curve. For details, See the Equ. 1 and Fig. 2.

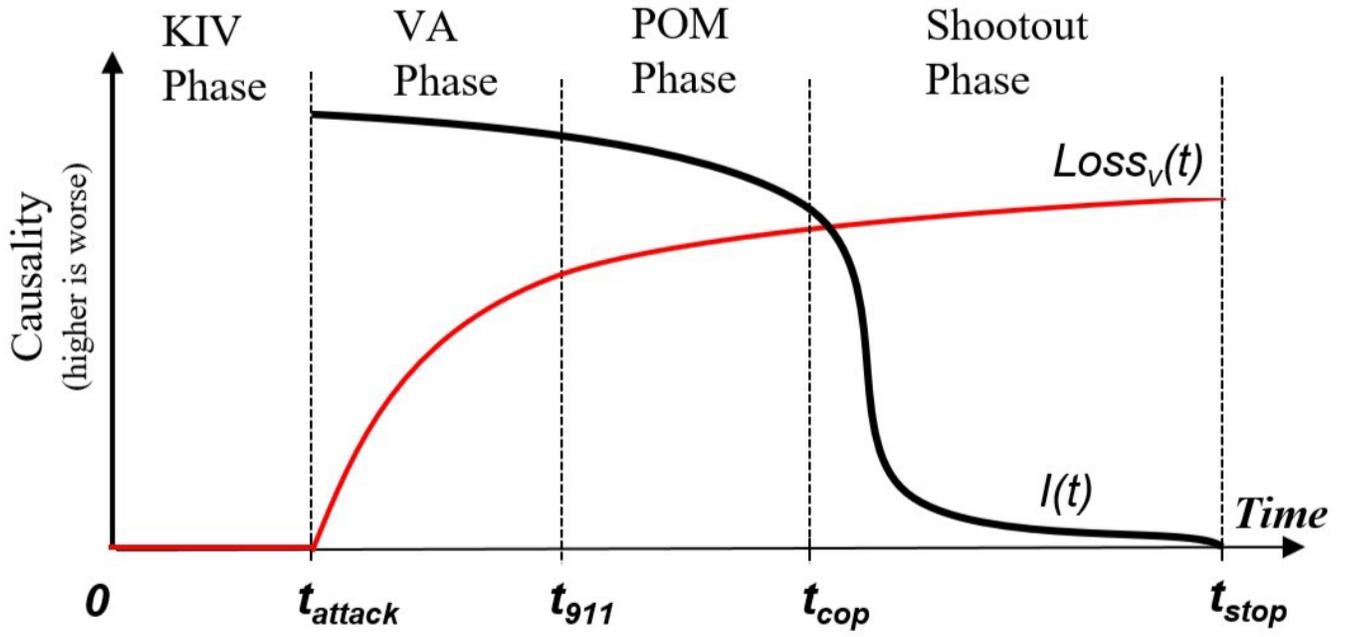

Fig. 2. Miller Curve explaining the relationship between accumulated human causalities, $Loss_v(t)$, and injure scale of innocents, $I(t)$.

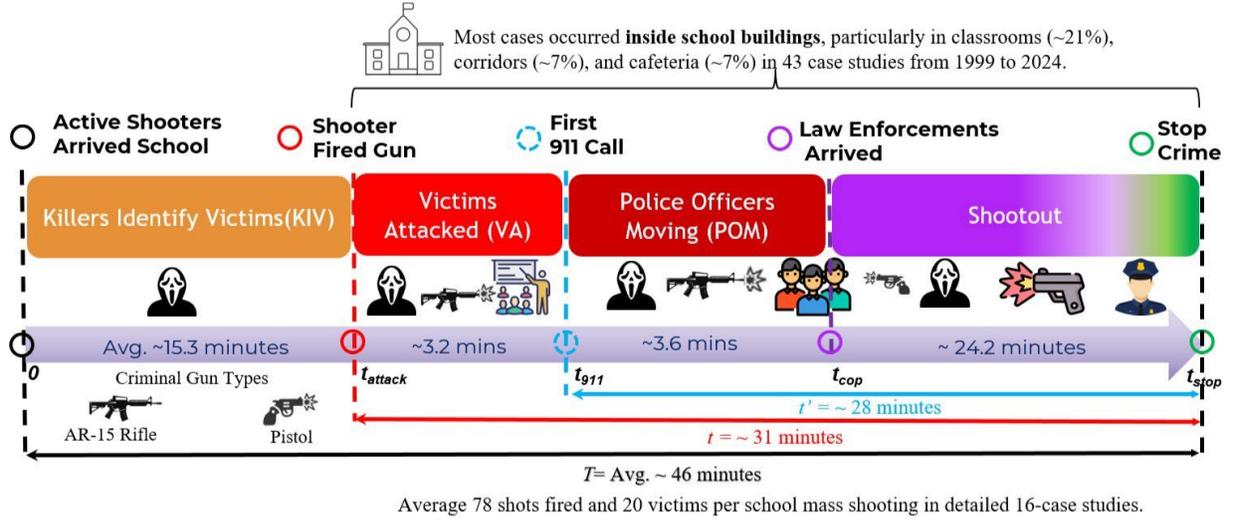

Fig. 3. Timeline of the 16-incident Mass School Shootings in the USA

$$\arg\min_b \max_t Loss_v(t) + Loss_m(t) =$$

$$\arg\min_b \max_t \sum_{i=1}^{n}\sum_{j=1}^{m} I(t,b,i)V_j(t) + \sum_{i=1}^{n} M(t,i) \quad (1)$$

$$\text{subject to } i \in n, j \in m, t \in T, \text{ and } b \in B$$

where

$Loss_v(t)$ and $Loss_m(.)$ are the loss functions of human causalities and material damage, respectively. $i$ is the killer $i$. $j$ is the victim $V_j$. $t$ is the timeline of the shooting. $b$ is the number of bullets. $I(.)$ is the injury scale of victim. $M(.)$ is the cost of material damage.

Law enforcement officers, $p$, have variable weapons, $w$. Compared to active shooters, we assume that the number of law enforcement and their weapons could be increased without limitation. However, the restriction of law enforcement officers is unanticipated costs, $U(.)$, when gun battles are in progress. For example, police officers accidentally shot their own people or innocent people during a shootout. Equ. 2 explains the strategy formula of law enforcements.

$$\arg\min_{u}\max_{t'} \sum_{p=1}^{+\infty}\sum_{i=1}^{n} S(t',w,p)K_i(t') - \sum_{i=1}^{n} U(t',i) \qquad (2)$$

$$\text{subject to } t' \in T,\ s,\ w \in +\infty$$

where

The killer is $K_i$. $t'$ is the time; $p$ is a law enforcement officer. $w$ is the weapon of law enforcement. $S(.)$ is the probability of stopping the crime. $U(.)$ is the unexpected causality.

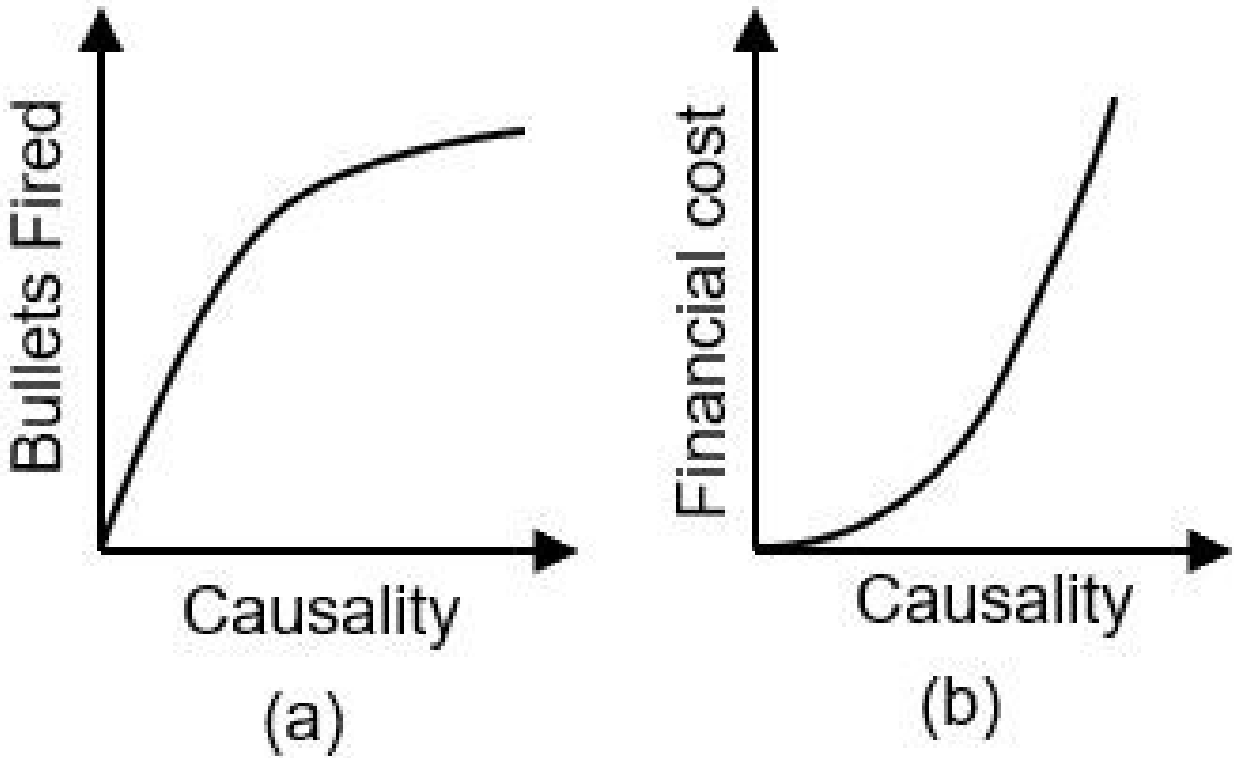

Fig. 4. The relationship in Bullet, Causality, and Financial Cost

## IV. PROBABILITY OF SHOOTING EVENTS IN SCHOOLS

There are a total of 553 events related to schools in the nation from 1999 to 2024 (or 26 years), including 510 (or 92.2%) school shootings events and 43 (or 7.8%) mass school shootings events. Detailed Information of the shootings in schools in the USA for 26 years, see Table I. Each tile represents a state in Fig. 1, with colors ranging from light purple to dark purple to indicate the frequency of incidents. For instance, California is marked in the darkest shade of purple, reflecting the highest number of school shootings at 49. Just one shade lighter, states like Texas, Florida, and North Carolina fall within the range of 29 to 3 shootings. In contrast, Vermont and Maine are left uncolored (gray), signifying that no school shootings have been reported in those states. The lighter blue hues represent states with fewer incidents, ranging from 1 to 9 shootings. This color coding effectively highlights the stark disparities in school shootings occurrences across the nation. The red bold color in the middle of the majority of states represents the events of mass school shootings from 1999 to 2024.

School shootings and mass school shootings have strong correlation. We choose the Pearson correlation coefficient and Pvalue to evaluate the numbers of school shootings and mass school shootings in each state, the Person correlation coefficient is 0.754, which means that it indicates a strong positive correlation. We also calculate a two-tailed P-value from two datasets. The P-value is $4.48 \times 10^{-7}$, which means that it considered to be very highly statistically significant.

In 2022, the U.S. had 134,960 schools in all types[15]. Despite the number of schools always change a little bit, we assume that school number, 134,960, is a constant. There are total 510 school shootings and 43 mass school shootings during 26 years (1999-2024), which means that the events of school shootings and mass school shootings are 19.62 events and 1.65 events per year, respectively. The yearly probability of school shootings is $1.45 \times 10^{-4}$ (or one in 6880) per school; the annual probability of mass school shootings is $1.23 \times 10^{-5}$ (or one in 81,604) per school.

Assume that a student spends 17 years in schools, including Kindergarten education, six years of elemental education, four years of Junior High School, and two years of Senior High School, and a 4-year college education. In the US, the student has a total 0.245% probability of encountering a school shooting (one in 408) and a total 0.021% probability of encountering a mass school shooting (one in 4801) for his/her 17-year education.

## V. ANALYSIS OF MASS SCHOOL SHOOTINGS

From 1999 to 2024, the tragic occurrence of 43 mass school shootings has left a profound impact on communities across the United States, with average 8 deaths and average 13 injured ( or total 21 casualties) per mass school shootings. the Columbine High School incident in Littleton, Colorado to the Covenant School in Nashville, Tennessee. For historical data of the mass school shootings in the USA, for 26 years, see Table V. The most common weapon is the AR-15 style rifle. .

### A. Analyzing Criminal Locations of Mass School Shootings

Most mass school shootings occur inside school buildings. For details, see Fig. 5. The chart depicts the frequency of incidents by location in 43 mass school shootings from 1999 to 2024. Classrooms (13 events, or 30.23%) are the most frequent locations where incidents occur, followed by hallways (9 events, or 20.93%) and outside (6 events, or 13.95%). Other areas, such as parking lots, entry doors, cafeterias, gyms, and offices, see fewer incidents, with the least frequent places being fields, libraries, bathrooms, and buses.

Classrooms and hallways are the most vulnerable locations for incidents, which suggests that preventive measures should be given particular attention. This may be because classrooms typically hold the most students, making them a target for active shooters seeking to cause maximum harm. Hallways are high-traffic areas, especially during transitions between classes, providing opportunities for attackers to target large groups of people. In contrast, areas with lower frequencies, like libraries and bathrooms, might see fewer incidents because they tend to have fewer people and provide less visibility for an attacker, making them less attractive targets. Enhance security and monitoring systems in classrooms and hallways, as these are the most frequent incident locations. Implement effective lockdown procedures, increase the visibility of security personnel, and consider advanced surveillance technology in these high-risk areas. Since these areas have high privacy requirements, implementing gunshot detection systems that do not infringe on personal privacy is also very important.

### B. Analyzing Criminal Timelines of Mass School Shootings

In this study, we collected 16 mass school shootings events providing detailed timelines. For details, see Fig. 3. First, a shooter arrives at the school and prepares to act, which takes an average of 15 minutes as the KIV (or Killers Identify Victims) Phase. Then, the Victims Attacks (or VA) Phase is defined as the time between an active shooter firing his/her guns and the first 911 call. It takes an average of almost three minutes. Third, the POM (Police Officers Moving) Phase indicates that law enforcement arrives at the scene of the crime after receiving the 911 call. The POM Stage takes an average of almost four minutes. Lastly, law enforcement takes an average of almost 24 minutes to stop criminal activity, defining as Shootout Phase. The average duration of criminal time is 31 minutes from VA Phase to Shootout Phase. According to 16 different cases of school shootings from Columbine to Nashville, there is an average of 20 innocent victims with 78 rounds fired by the shooter. The average criminal time is 31 minutes with almost 20 casualties. Thus, there is 0.639 casualties per minute, which means that almost two causalities every three minutes.

*C. Analyzing Causalities of Mass School Shootings*

The mass school shootings always bring causalities. In this research, we study five factors that affect the casualties in the 16 mass school shootings. The Pearson correlation coefficient, $r$, and $P$ values are selected to evaluate the correlation coefficient. For details, see Table VI. For the Police Station Distance and Causality, $r$ is -0.342, showing a modern negative correlation with 99% confidence. If a school is far from a police station, the number of victims may increase. For the Hospital Distance and Causality, $r$ is -0.162, showing a weak negative correlation with 99% confidence.

For the bullets fired and Causality, $r$ is 0.592, indicating a strong positive correlation with 99% confidence ($p < 0.01$). We also identify that the number of bullets fired is increased, and the number of casualties is slowly reduced. When a killer fires bullets, people may adopt the "run, hide, fight" policies, resulting in fewer deaths. Unfortunately, the financial costs increase very rapidly as the number of victims increases. Financial costs include both direct and indirect costs [16] [17] [18] [19]. For details, see Fig. 4.

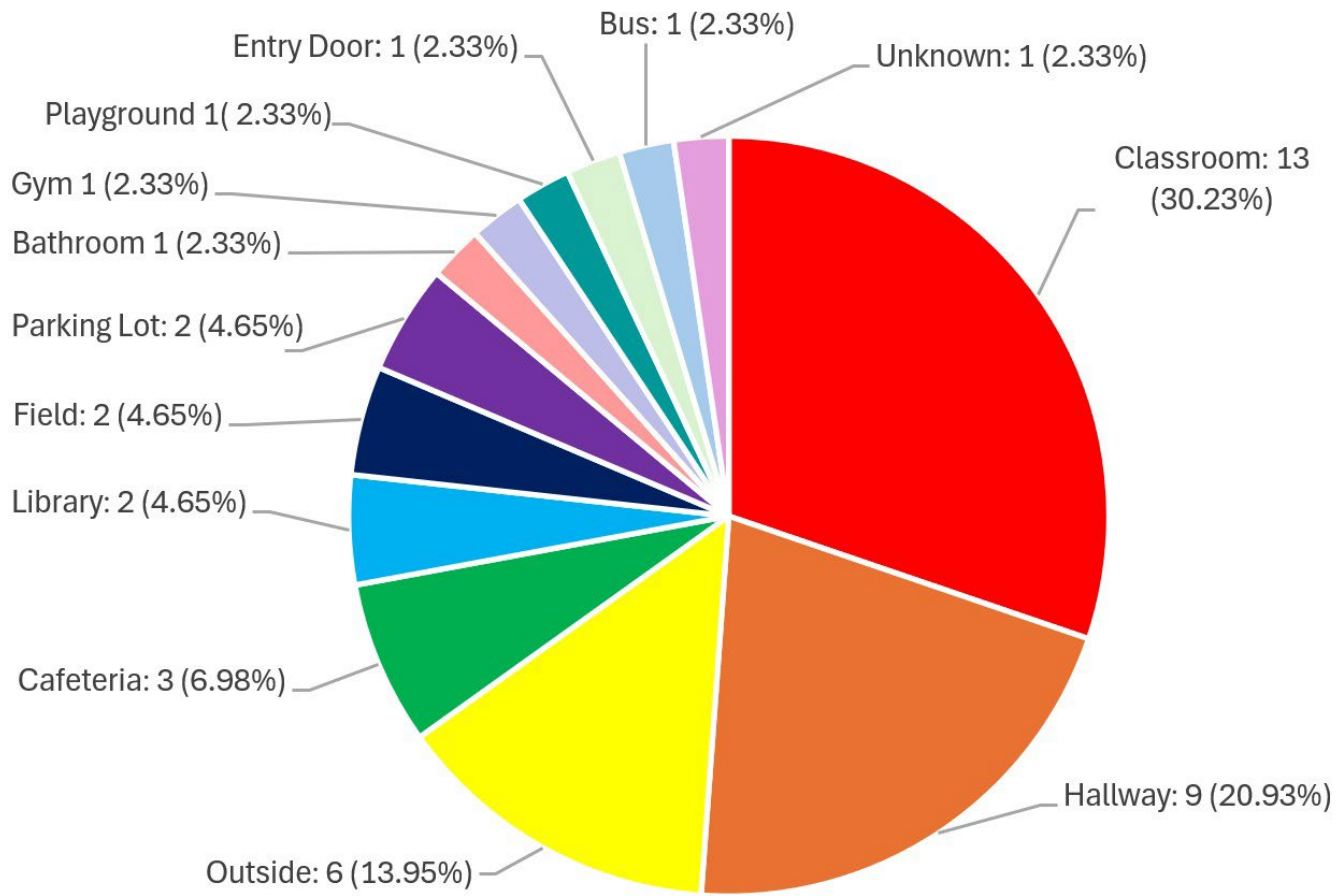

Fig. 5. Analysis of Locations in Mass School Shootings

TABLE I
HISTORICAL INCIDENTS AND CASUALTIES OF THE MASS SCHOOL SHOOTINGS IN THE USA FROM 1999 TO 2024

| Year | 1999 | 2000 | 2001 | 2002 | 2003 | 2004 | 2005 | 2006 | 2007 | 2008 | 2009 | 2010 | 2011 | 2012 |
|---|---|---|---|---|---|---|---|---|---|---|---|---|---|---|
| Events | 3 | 0 | 2 | 0 | 1 | 0 | 1 | 1 | 2 | 1 | 0 | 0 | 0 | 3 |
| Innocents Injured | 35 | 0 | 18 | 0 | 4 | 0 | 5 | 5 | 36 | 16 | 0 | 0 | 0 | 6 |
| Innocents Killed | 13 | 0 | 2 | 0 | 1 | 0 | 9 | 5 | 32 | 5 | 0 | 0 | 0 | 36 |
| Total Casualty | 48 | 0 | 20 | 0 | 5 | 0 | 14 | 10 | 68 | 21 | 0 | 0 | 0 | 42 |
| Year | 2013 | 2014 | 2015 | 2016 | 2017 | 2018 | 2019 | 2020 | 2021 | 2022 | 2023 | 2024 | Total | Avg |
| Events | 1 | 2 | 1 | 2 | 1 | 4 | 4 | 0 | 1 | 5 | 6 | 2 | 43 | 1.65 |
| Innocents Injured | 3 | 17 | 7 | 8 | 3 | 53 | 31 | 0 | 7 | 36 | 20 | 13 | 323 | 12.42 |
| Innocents Killed | 5 | 10 | 9 | 0 | 1 | 29 | 3 | 0 | 4 | 26 | 17 | 4 | 211 | 8.12 |
| Total Casualty | 8 | 27 | 16 | 8 | 4 | 82 | 34 | 0 | 11 | 62 | 37 | 17 | 534 | 20.54 |

*D. Predicting 5-year Mass School Shootings with Machine Learning Models*

Mass school shootings are rare, but these events unfortunately increased during the COVID-19 pandemic. The start date of the COVID-19 public health emergency (COVID-19 pandemic for short) was January 31, 2020, and the end date was May 11, 2023. Existing research studies have found that with the onset of the COVID-19 pandemic, there has been a sharp increase in mass shootings and gun violence in the United States.[20] [21] [22]. In this study, we define the COVID-19 period as the four years from 2020 to 2023. There were 29 mass school shooting events between 1999 and 2019 (or 21 years). Thus, the average case numbers are 1.38 yearly events before the COVID-19 pandemic. Fortunately, there were zero events in six years, including 2000, 2002, 2004, 2009, 2010, and 2011. If we exclude these six years, the average number of mass school shootings is 1.93 events per year. The number of mass school shootings is two, five, and six in 2021, 2022, and 2023, respectively. The average number of cases is 3.25 events per year during COVID-19, 2.36 times higher than the average number between 1999 and 2019. Note that In 2020, there were zero cases of mass school shootings because the quarantine and isolation policy stopped face-to-face instruction in schools. The average number of cases is 4.33 events per year during the pandemic period if we exclude 2020. Compared to 1.93 events per year, the annual 4.33 events during the pandemic period are 2.24 times greater than its peers.

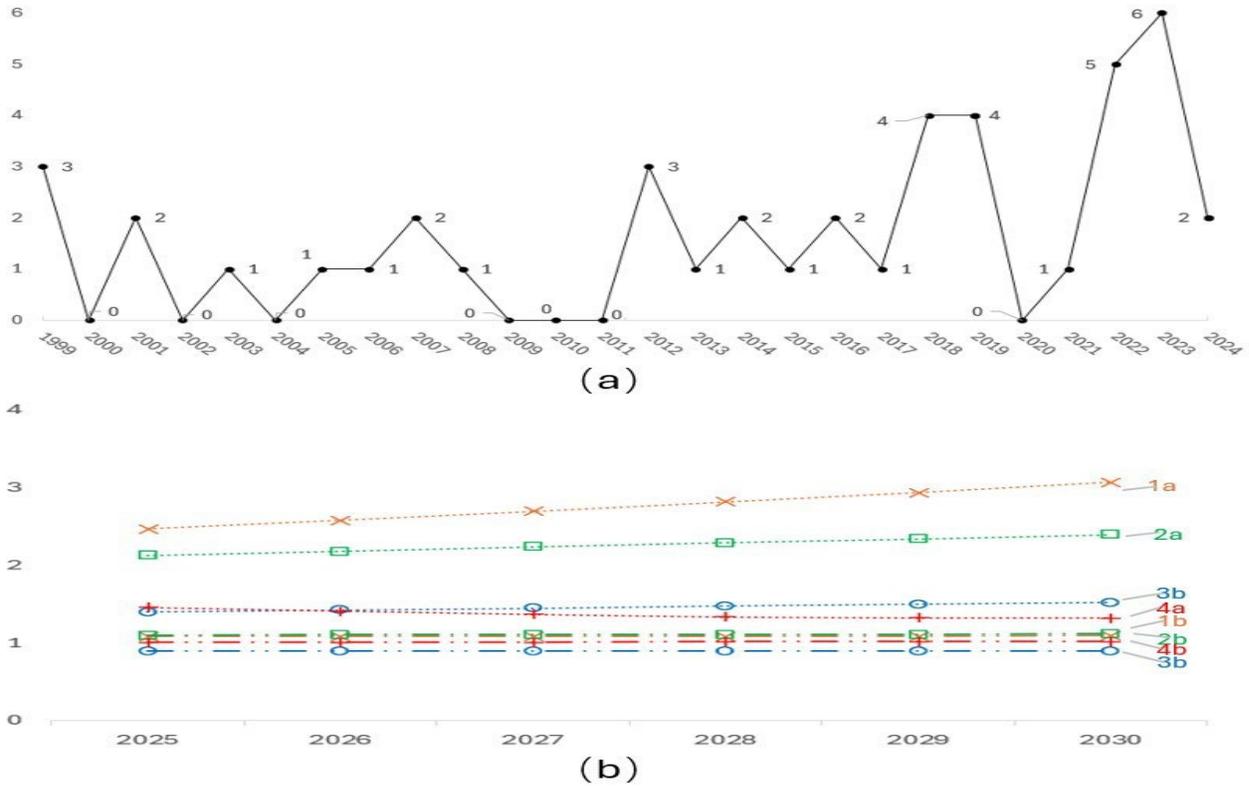

Fig. 6. Incidents in Mass School Shootings in the USA from 1999 to 2030 Note that (a) is the existing events; (b) is the forecast events

However, it is essential to note that the increase observed during the pandemic could be influenced by other factors, such as social unrest, economic instability, or changes in gun ownership and policy during this time. This complexity highlights the need for a more nuanced analysis to disentangle the specific contributions of the pandemic from other overlapping variables. Therefore, identifying the appropriate models and datasets for prediction is crucial to capturing the underlying patterns and providing accurate insights into the factors contributing to these events.

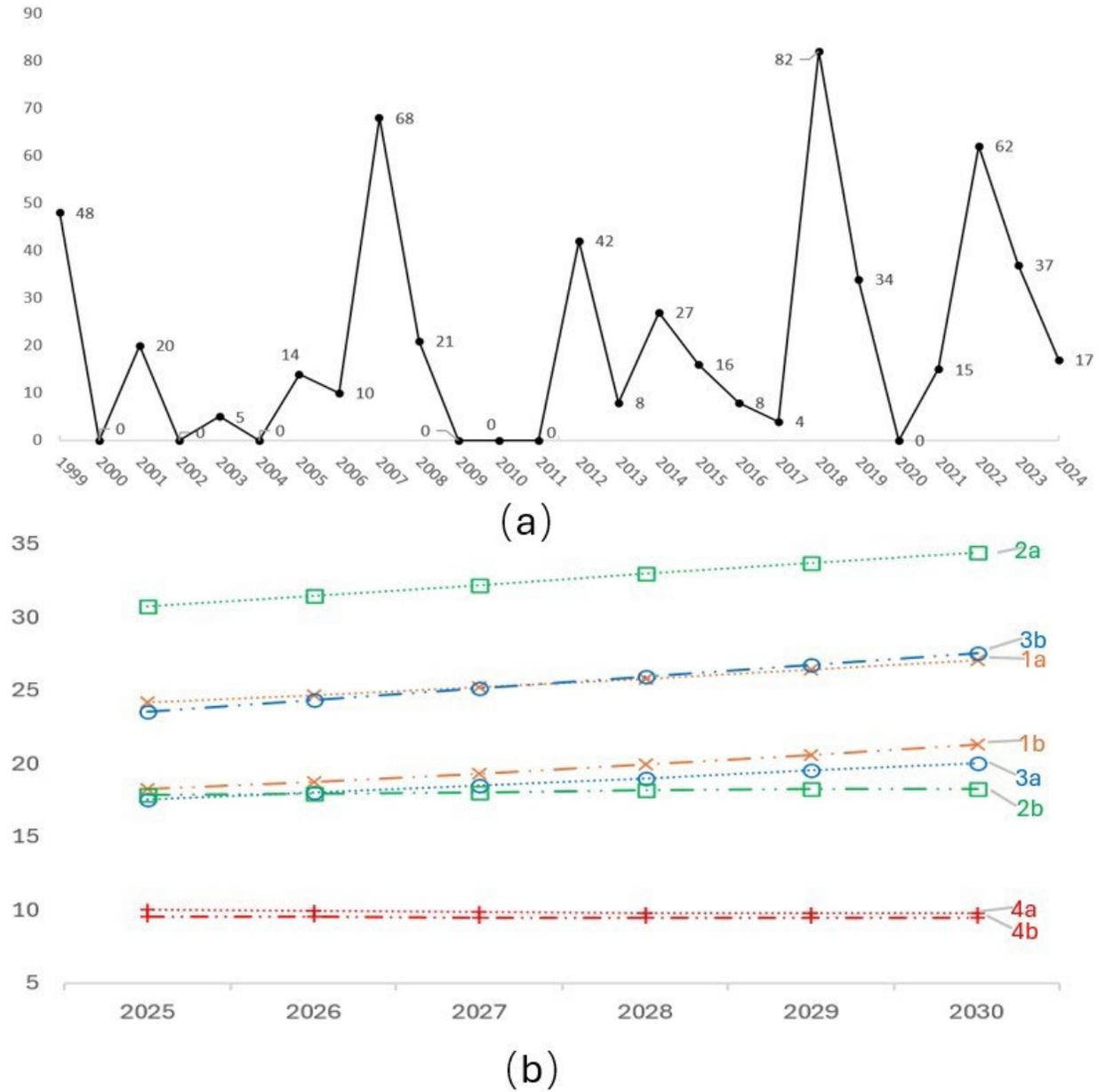

Fig. 7. The number of Casualties in Mass School Shootings in the USA from 1999 to 2030 Note that (a) is the existing causalities; (b) is the forecast causalities

We utilized four prediction machine learning models, including ZIP (Zero-Inflated Poisson), designed for datasets with excessive zeros, combining Poisson regression and binary logistic regression to handle zero-inflated, count-based data. Linear

Regression: Assumes a linear relationship between variables, minimizing squared residuals for simplicity. SVR with Linear Kernel: Maintains linearity while mapping input features to a high-dimensional space suitable for approximating linear relationships. SVR with RBF Kernel: Captures complex non-linear relationships using a radial basis function, offering flexibility for intricate patterns.

We used two training datasets for predictions: the dataset with COVID-19 pandemic data, which includes data from 20202023, and the dataset without COVID-19 pandemic data, which excludes data from 2020-2023. We used both datasets to predict events for the future 5-year period (2025-2030) mass shooting events.

When prioritizing the Mean Squared Error (MSE) and Mean Absolute Error (MAE), linear regression outperformed other models. When prioritizing Mean Absolute Percentage Error (MAPE), SVR with Linear Kernel performed best for both datasets. Best Dataset: Overall, the dataset without COVID-19 data performed better, consistently producing lower MSE, MAPE, and MAE values. This finding indicates that removing pandemic-related data allows the model to achieve more accurate predictions.

The final results, visualized in Figs. 6-7, illustrate the predictions of incidents count and casualties count. These predictions are detailed in Tables II - III. For casualties count, as detailed in Tables II and III, predictions suggest a relatively stable trend from 2025 onward, fluctuating between 1 and 3 cases per year. These results highlight a potential decline and stabilization in both incidents and casualties, suggesting improved safety in future years.

## VI. CASE STUDY

### A. Apalachee High School

On September 4, 2024, Apalachee High School, a public school, in Winder, Georgia, had endured a mass shooting incident where 4 people were shot and 9 left injured. The 14-year-old student named Colt Gray left his classroom at 9:45 and was denied reentry when a student spotted an AR-15 style rifle wrapped around his torso. He begin shooting about 15 rounds at around 10:18 a.m. and the authorities were notified shortly at 10:20. The police officers took approximately two minutes to come and stopped the shooting incident at 12:25 p.m.

### B. The Covenant School

On March 27, 2023, The Covenant School, a private Christian School in Nashville, Tennessee, was shot by a transgender man named Aiden (Audrey Elizabeth) Hale and killed 6 people. Hale arrived at 9:54 a.m. at the parking lot and spent time sending Instagram messages to his friend. He then shot through the glass door at 10:11 a.m. and police received a phone call at 10:13 a.m. At 10:23 a.m. officers entered the building and at 10:27 the officers shot Hale. Hale fired more than 150 rounds with his AR-15, Kel-Tec SUB-2000 carbine, and a handgun. The perpetrator was 28-year-old with no criminal record and was a former student. Hale was under care for an emotional disorder and had also legally purchased seven firearms.

### C. Santa Fe High School

On May 18, 2018, 10 people were shot at Santa Fe High School in Santa Fe, Texas. Dimitrios Pagourtzis, a 17-year-old who was a student at the school, arrived around 7:00 a.m and began shooting at 7:32 a.m. Pagourtizs shot eight students and two teachers. Around 7:40 a.m. the first 911 was reported and police arrived five minutes later. He argued with the police officers and threatened to shoot, but surrendered after being injured. At 8:09 a.m. the shooter was arrested. The shooter shot 18 rounds of ammunition.

### D. Oxford High School

On November 30, 2021, a 15-year-old named Ethan Crumbly shot the Oxford High School in Oxford Township, Michigan. Crumbly killed 4 people with around 30 rounds of ammunition He was seen entering the bathroom with a book sack at 12:50 p.m. Then, he was seen with a gun and began shooting at 12:51 p.m. Police officers received their first 911 call at 12:52 p.m. and arrived at 12:56 p.m. Crumbly was arrested around 1:04. His parents were also charged with involuntary manslaughter for failing to secure the gun and negligence for not noticing warning signs by Crumbly.

### E. Marjory Stoneman Douglas High School

On February 14, 2018, 17 people were shot at Marjory Stoneman Douglas High School in Parkland, Florida. The 19-year-old former student, Nikolas Cruz, arrived at the school at 2:19 p.m. in an Uber and was armed with an AR-15 rifle. His first shot was around 2:21 p.m. The first 911 call was around 2:22 p.m. and police arrived around 2:23 p.m. The shooter was caught at 2:50 p.m.

### F. Umpqua Community College

On October 1, 2015, 9 people were shot to death by a 26-year-old student who was attending Umpqua College. Chris Harper-Mercer shot 40 rounds killing a professor and eight students and eventually shot himself. Harper-Mercer was first seen at the location at 10:39 a.m and shot his first round at 10:32 a.m. The first 911 was made at 10:38 a.m. He asked the students for their religion before shooting and if they were Christians he would shoot them. Police officers had arrived at 10:44 a.m. and began shooting the perpetrator. Harper-Mercer had been wounded and later shot himself at 10:48 am.

*G. West Nickel Mines Amish School*

On October 22, 2006 a 32-year-old Charles Carl Roberts IV killed 5 little girls at the West Nickel Mines School in Bart Township, Pennsylvania. Roberts was seen at the front of the school around 10:25 a.m. He was seen with a Springfield 9 mm handgun and a 911 call was made at 10:36 a.m. Police officers arrived at the area around 10:42 a.m. By 11:00 a.m. a large crowd began to form outside the school, where Roberts had the girls hostage. Around 11:05 a.m. began to shoot the girls and by 11:07 he shot himself to death.

*H. Columbine High School*

On April 20, 1999 at Columbine High School in Columbine, Colorado, two perpetrators Eric Harris and Dylan Klebold killed 13 people by shooting and bombing with a Stevens 311D shotgun. Around 11:15 they plan to detonate their explosives, which failed so at 11:19 Klebold and Harris began shooting. At 11:22 a.m. 911 was called and police arrived 2 minutes later. As the police began to chase them, they later shot themselves around 12:08 p.m.

## VII. DATA COLLECTING METHOD

We identify incidents of school shootings on the Gun Violence Archive website (https://www.gunviolencearchive.org/). We then cross-checked the data sets from various news sources. The distances between schools to the nearest police station and hospital are measured through Google map.

## VIII. CONCLUSION

Gun violence in the United States has raised numerous concerns about the safety of schools. Existing research that includes statistical analysis usually focuses on public mass shootings or just shooting incidents that have occurred in the past, and there are hardly any articles focusing on mass school shootings. This makes schools more vulnerable to mass shootings in the future. In this paper, the authors designed the model of game theory to explain the strategies of active killers and law enforcement. In addition, the authors evaluated the probability rate, analyzed timelines of mass school shootings, and predicated 5-year incidents of mass school shootings. By interpreting these data and conducting various data analysis, this will ultimately help law enforcement better prepare.

## ACKNOWLEDGMENT

Thank Mr. Brian Miller, Dr. Yu Ouyang, Dr. Bree Alexander, and Mr. Abhishek Singh for their involvement.

TABLE II
PREDICATED INCIDENTS OF THE MASS SCHOOL SHOOTINGS IN THE USA FROM 2025 TO 2030

| Model ID | Model Name | 2025 | 2026 | 2027 | 2028 | 2029 | 2030 | MSE | MAE | Training Data |
|---|---|---|---|---|---|---|---|---|---|---|
| 1a | Zero-Inflated Poisson | 2.47 | 2.58 | 2.70 | 2.82 | 2.94 | 3.07 | 5.07 | 1.85 | 1999-2024 |
| 1b | Zero-Inflated Poisson | 1.09 | 1.09 | 1.09 | 1.09 | 1.09 | 1.10 | 3.82 | 1.59 | 1999-2021 and 2024 |
| 2a | Linear Regression | 2.13 | 2.18 | 2.24 | 2.29 | 2.34 | 2.40 | 5.54 | 1.87 | 1999-2024 |
| 2b | Linear Regression | 1.10 | 1.11 | 1.11 | 1.11 | 1.11 | 1.12 | 3.80 | 1.58 | 1999-2021 and 2024 |
| 3a | SVR Linear | 1.40 | 1.43 | 1.45 | 1.48 | 1.50 | 1.52 | 7.53 | 2.28 | 1999-2024 |
| 3b | SVR Linear | 0.90 | 0.90 | 0.90 | 0.90 | 0.90 | 0.90 | 4.25 | 1.66 | 1999-2021 and 2024 |
| 4a | SVR RBF | 1.46 | 1.41 | 1.37 | 1.34 | 1.33 | 1.32 | 6.42 | 2.06 | 1999-2024 |
| 4b | SVR RBF | 1.01 | 1.01 | 1.01 | 1.01 | 1.02 | 1.02 | 3.86 | 1.58 | 1999-2021 and 2024 |
| - | Average | 1.44 | 1.46 | 1.48 | 1.50 | 1.53 | 1.56 | - | - | - |

TABLE III
PREDICATED CASUALTIES OF THE MASS SCHOOL SHOOTINGS IN THE USA FROM 2025 TO 2030

| Model ID | Model Name | 2025 | 2026 | 2027 | 2028 | 2029 | 2030 | MSE | MAE | Training Data |
|---|---|---|---|---|---|---|---|---|---|---|
| 1a | Zero-Inflated Poisson | 24.18 | 24.70 | 25.25 | 25.82 | 26.42 | 27.05 | 157.51 | 10.53 | 1999-2024 |
| 1b | Zero-Inflated Poisson | 18.25 | 18.78 | 19.34 | 19.95 | 20.60 | 21.30 | 389.01 | 14.38 | 1999-2021 and 2024 |
| 2a | Linear Regression | 30.77 | 31.50 | 32.24 | 32.98 | 33.71 | 34.44 | 135.80 | 10.10 | 1999-2024 |
| 2b | Linear Regression | 17.91 | 18.00 | 18.08 | 18.17 | 18.25 | 18.32 | 401.67 | 14.88 | 1999-2021 and 2024 |
| 3a | SVR Linear | 17.53 | 18.04 | 18.53 | 19.03 | 19.54 | 20.03 | 168.51 | 10.20 | 1999-2024 |
| 3b | SVR Linear | 23.60 | 24.40 | 25.19 | 25.99 | 26.78 | 27.57 | 341.69 | 13.89 | 1999-2021 and 2024 |
| 4a | SVR RBF | 10.03 | 9.93 | 9.86 | 9.81 | 9.78 | 9.77 | 200.27 | 10.47 | 1999-2024 |
| 4b | SVR RBF | 9.54 | 9.52 | 9.51 | 9.50 | 9.50 | 9.50 | 350.21 | 12.29 | 1999-2021 and 2024 |
| - | Average | 18.98 | 19.36 | 19.75 | 20.16 | 20.57 | 21.00 | - | - | - |

TABLE IV

CAUSALITY, BULLET, TIMELINE, AND DISTANCES OF THE 16-INCIDENT MASS SCHOOL SHOOTINGS IN THE USA

| ID | School Name, City, State | Casualty | Bullets Fired | KIV Phase (mins) | VA Phase (mins) | POM Phase (mins) | Shootout Phase (mins) | Distance to Police Station | Distance to Hospital | Crime Time (mins) |
|---|---|---|---|---|---|---|---|---|---|---|
| 1 | Apalachee High School, Winder, Georgia | 13 | 15 | 63 | 2 | 3 | 7 | 9.2 km (5.7mi) | 3.2 km (2.0mi) | 12 |
| 2 | The Covenant School, Nashville, Tennessee | 7 | 150 | 17 | 2 | 10 | 4 | 7.7 km (4.8mi) | 6.9 km (4.3mi) | 16 |
| 3 | Michigan State University, East Lansing, Michigan | 8 | 18 | 53 | 6 | 2 | 209 | 2.6 km (1.6mi) | 2.1 km (1.3mi) | 217 |
| 4 | Oxford High School, Oxford, Michigan | 11 | 30 | 1 | 1 | 4 | 8 | 2.9 km (1.8mi) | 10.3 km (6.4mi) | 13 |
| 5 | Santa Fe High School, Santa Fe, Texas | 23 | 18 | 32 | 8 | 5 | 24 | 6.3 km (3.9mi) | 9.3 km (5.8mi) | 37 |
| 6 | Marjory Stoneman Douglas High School, Parkland, Florida | 34 | 139 | 2 | 1 | 1 | 27 | 2.3 km (1.4mi) | 5.3 km (3.3mi) | 29 |
| 7 | Umpqua Community College, Roseburg, Oregon | 16 | 40 | 2 | 6 | 6 | 4 | 1.8 km (1.1mi) | 8.4 km (5.2mi) | 16 |
| 8 | Marysville-Pilchuck High School, Marysville, Washington | 7 | 8 | 9 | 0 | 1 | 3 | 6.0 km (3.7mi) | 8.0 km (5.0mi) | 4 |
| 9 | University of California, Santa Barbara, California | 20 | 55 | 3 | 9 | 3 | 7 | 1.6 km (1.0mi) | 4.3 km (2.7mi) | 19 |
| 10 | Sandy Hook Elementary School, Newton, Connecticut | 27 | 154 | 0 | 5 | 2 | 3 | 8.2 km (5.1mi) | 4.5 km (2.8mi) | 10 |
| 11 | Oikos University, Oakland, California | 10 | 10 | 3 | 3 | 6 | 10 | 3.1 km (1.9mi) | 3.7 km (2.3mi) | 19 |
| 12 | Northern Illinois University, Dekalb, Illinois | 21 | 50 | 5 | 1 | 0 | 5 | 1.1 km (0.7mi) | 5.3 km (3.3mi) | 6 |
| 13 | Virginia Tech University, Blacksburg, Virginia | 58 | 174 | 5 | 2 | 3 | 6 | 0.5 km (0.3mi) | 6.1 km (3.8mi) | 11 |
| 14 | West Nickel Mines Amish School, Nickel Mines, Pennsylvania | 11 | 13 | 40 | 0 | 6 | 25 | 10.8 km (6.7mi) | 24.6 km (15.3mi) | 31 |
| 15 | Red Lake High School, Red Lake, Minnesota | 14 | 59 | 6 | 2 | 4 | 1 | 1.3 km (0.8mi) | 1.4 km (0.9mi) | 7 |
| 16 | Columbine High School, Littleton, Colorado | 37 | 188 | 4 | 3 | 2 | 44 | 4.3 km (2.7mi) | 3.7 km (2.3mi) | 49 |
|  | Average | 19.8 | 78.3 | 15.3 | 3.2 | 3.6 | 24.2 | 4.4 km (2.7mi) | 6.7 km (4.2mi) | 31 |

TABLE V: Historical Data of the Mass School Shootings in the USA from 1999 to 2024

| ID | School Name, City, State | School Type | Incident Date | Innocents killed | Innocents Injured | Bullets | Shooter Arrived | Shooter Fired Gun | 911 Call | Law Enforcement Arrived | Stop Crime | Shooter's Firearms |
|---|---|---|---|---|---|---|---|---|---|---|---|---|
| 1 | Apalachee High School, Winder, Georgia | public | 09/04/2024 | 4 | 9 | 15 | 9:50 AM | 10:18 AM | 10:20 AM | 10:23 AM | 10:30 AM | AR-15 |
| 2 | The Covenant School, Nashville, Tennessee | private | 03/27/2023 | 6 | 1 | 150 | 9:54 AM | 10:11 AM | 10:13 AM | 10:23 AM | 10:27 AM | AR-15 |
| 3 | Michigan State University, East Lansing, Michigan | public | 02/13/2023 | 3 | 5 | 18 | 7:19 PM | 8:12 PM | 8:18 PM | 8:27 PM | 11:49 PM | 9 mm handgun |
| 4 | Oxford High School, Oxford, Michigan | public | 11/30/2021 | 4 | 7 | 30 | 12:50 PM | 12:51 PM | 12:52 PM | 12:56 PM | 1:04 PM | SIG Pro 2022 |
| 5 | Santa Fe High School, Santa Fe, Texas | public | 05/18/2018 | 10 | 13 | 18 | 7:00 AM | 7:32 AM | 7:40 AM | 7:45 AM | 8:09 AM | Remington 870 |
| 6 | Marjory Stoneman Douglas High School, Parkland, Florida | public | 02/14/2018 | 17 | 17 | 139 | 2:19 PM | 2:21 PM | 2:22 PM | 2:23 PM | 2:50 PM | AR-15 |
| 7 | Umpqua Community College, Roseburg, Oregon | public | 10/01/2015 | 9 | 7 | 40 | 10:30 AM | 10:32 AM | 10:38 AM | 10:44 AM | 10:48 AM | Glock 19 |
| 8 | Marysville-Pilchuck High School, Marysville, Washington | public | 10/14/2014 | 4 | 3 | 8 | 10:30 AM | 10:39 AM | 10:39 AM | 10:40 AM | 10:43 AM | Beretta Px4 |
| 9 | University of California, Santa Barbara, California | public | 05/23/2014 | 6 | 14 | 55 | 9:15 PM | 9:18 PM | 9:27 PM | 9:30 PM | 9:37 PM | Glock 34 |
| 10 | Sandy Hook Elementary School, Newton, Connecticut | public | 12/14/2012 | 27 | 0 | 154 | 9:30 AM | 9:30 AM | 9:35 AM | 9:37 AM | 9:40 AM | AR-15 |
| 11 | Oikos University, Oakland, California | private | 04/02/2012 | 7 | 3 | 10 | 10:27 AM | 10:30 AM | 10:33 AM | 10:39 AM | 10:49 AM | .45 ACP pistol |
| 12 | Northern Illinois University, Dekalb, Illinois | public | 02/14/2008 | 5 | 16 | 50 | 3:00 PM | 3:05 PM | 3:06 PM | 3:06 PM | 3:11 PM | Remington Model 11-48 |
| 13 | Virginia Tech University, Blacksburg, Virginia | public | 04/16/2007 | 32 | 26 | 174 | 9:15 AM | 9:40 AM | 9:42 AM | 9:45 AM | 9:51 AM | Glock 19, Walther P22 |
| 14 | West Nickel Mines Amish School, Nickel Mines, Pennsylvania | private | 10/02/2006 | 5 | 6 | 13 | 10:25 AM | 11:05 AM | 10:36 AM | 10:42 AM | 11:07 AM | Springfield XD 9mm |
| 15 | Red Lake High School, Red Lake, Minnesota | public | 03/21/2005 | 9 | 5 | 59 | 2:45 PM | 2:51 PM | 2:53 PM | 2:57 PM | 2:58 PM | Ruger MK II, Glock 22 |
| 16 | Columbine High School, Littleton, Colorado | public | 04/20/1999 | 13 | 24 | 188 | 11:15 AM | 11:19 AM | 11:22 AM | 11:24 AM | 12:08 PM | Stevens 311D |
| 17 | Mays High School, Atlanta, Georgia | public | 02/14/2024 | 0 | 4 | - | - | 4:00 PM | - | - | - | AR-15 |
| 18 | Huguenot High School, Richmond, Virginia | public | 06/06/2023 | 5 | 2 | - | - | - | - | - | - | AR-15 |
| 19 | University of Nevada, Las Vegas, Nevada | public | 12/06/2023 | 3 | 3 | - | 11:30 AM | - | 10:45 AM | - | 11:55 AM | 9 mm handgun |
| 20 | Moran State University, Baltimore, Maryland | public | 10/03/2023 | 0 | 5 | - | - | 9:27 AM | - | - | - | - |
| 21 | Westinghouse Academy, Pittsburgh, Pennsylvania | public | 02/14/2023 | 0 | 4 | 10 | - | 2:23 PM | - | - | - | - |
| 22 | University of Virginia, Charlottesville, Virginia | public | 11/13/2022 | 3 | 2 | - | - | 10:15 PM | - | - | - | handgun |
| 23 | Central Visual and Performing Arts High School, St. Louis, Missouri | public | 10/24/2022 | 2 | 7 | - | - | - | 9:11 AM | 9:15 AM | 9:25 AM | AR-15 |
| 24 | Rudsdale High School, Oakland, California | public | 09/28/2022 | 1 | 5 | 30 | - | 12:40 PM | 12:45 PM | - | 3:00 PM | handgun |
| 25 | Robb Elementary School, Uvalde, Texas | public | 05/24/2022 | 21 | 17 | 164 | 11:28 AM | 11:28 AM | 11:29 AM | 11:35 AM | 12:50 PM | AR-15 |

TABLE V: Historical Data of the Mass School Shootings in the USA from 1999 to 2024

| ID | School Name, City, State | School Type | Incident Date | Innocents killed | Innocents Injured | Bullets | Shooter Arrived | Shooter Fired Gun | 911 Call | Law Enforcement Arrived | Stop Crime | Shooter's Firearms |
|---|---|---|---|---|---|---|---|---|---|---|---|---|
| 26 | Edmund Burke School, Washington, D.C. | private | 04/22/2022 | 0 | 4 | 239 | - | - | - | - | - | automatic rifle |
| 27 | Saugus High School, Santa Clarita, California | public | 11/14/2019 | 2 | 3 | - | - | 7:38 AM | - | - | - | .45 caliber |
| 28 | STEM School Highlands Ranch, Highlands Ranch, Colorado | public | 05/07/2019 | 1 | 8 | - | - | 1:53 PM | - | - | - | .45 caliber |
| 29 | Ladd-Peebles Stadium, Mobile, Alabama | public | 09/30/2019 | 0 | 10 | - | - | - | - | - | - | - |
| 30 | Wynbrooke Elementary Theme School, DeKalb County, Georgia | public | 04/25/2019 | 0 | 10 | - | - | - | - | - | - | pellet gun |
| 31 | Marshall County High School, Draffenville, Kentucky | public | 01/23/2018 | 2 | 18 | - | - | 7:57 AM | - | - | 8:06 AM | Ruger handgun |
| 32 | Salvador B. Castro Middle School, Los Angeles, California | public | 02/01/2018 | 0 | 5 | - | - | 8:55 AM | - | - | - | handgun |
| 33 | Freeman High School, Rockford, Washington | public | 09/13/2017 | 1 | 3 | - | - | - | - | - | - | AR-15 |
| 34 | June Jordan High School for Equity, San Francisco, California | public | 10/18/2016 | 0 | 4 | - | - | 3:15 PM | - | - | 4:00 PM | pistol |
| 35 | Madison High School Middletown, Ohio | public | 02/29/2016 | 0 | 4 | 6 | - | - | - | - | - | .38 caliber |
| 36 | Santa Monica College, Santa Monica, California | public | 06/07/2013 | 5 | 3 | - | - | - | 11:52 AM | - | 12:05 PM | .38 caliber |
| 37 | Chardon High School, Chardon, Ohio | public | 02/27/2012 | 3 | 3 | - | - | 7:30 AM | - | - | - | .22 caliber handgun |
| 38 | Springwater Trail High School, Gresham, Oregon | public | 04/10/2007 | 0 | 10 | - | - | - | - | - | - | .270 Winchester rifle |
| 39 | John McDonogh High School, New Orleans, Louisiana | public | 04/14/2003 | 1 | 4 | 20 | 10:30 AM | - | - | - | - | AK-47 |
| 40 | Santana High School, Santee, California | public | 03/05/2001 | 2 | 13 | - | - | 9:20 AM | - | - | - | .22-caliber rifle |
| 41 | Granite Hills High School, El Cajon, California | public | 03/22/2001 | 0 | 5 | - | - | 12:55 PM | - | - | - | pistol |
| 42 | Heritage High School, Conyers, Georgia | public | 05/20/1999 | 0 | 6 | - | - | - | - | - | - | .22-caliber rifle |
| 43 | Fort Gibson Middle School, Gibson, Oklahoma | public | 12/06/1999 | 0 | 5 | - | - | 7:45 AM | - | - | - | 9 mm pistol |

TABLE VI

CORRELATION RELATIONSHIP BETWEEN CAUSALITY AND 7 FACTORS

| Statistical Method | Bullets Fired | KIV Phase | VA Phase | POM Phase | Shootout Phase | Distance to Police Station | Distance to Hospital | Crime Time |
|---|---|---|---|---|---|---|---|---|
| Pearson Correlation, r | 0.592 | -0.341 | 0.041 | -0.354 | -0.148 | -0.342 | -0.162 | -0.162 |
| P-Value(2-tailed), p | 1.5E-03 | 5.3E-01 | 2.4E-04 | 5.5E-04 | 7.5E-01 | 9.9E-04 | 4.2E-03 | 4.3E-01 |